\documentclass[aps,prl,twocolumn,superscriptaddress,showpacs,showkeys,
notitlepage]{revtex4-1}

\usepackage{times,amsmath,amsfonts,amssymb,mathrsfs,graphics,graphicx,color,
comment,bm}
\usepackage[next]{inputenc}
\usepackage[dvips]{epsfig}
\usepackage[pdfstartview=FitH]{hyperref}
\usepackage{natbib}
\usepackage{pdfsync}
\usepackage{appendix}
\begin{document}
\title{$\mu$SR study of noncentrosymmetric superconductor PbTaSe$_2$}

\author{M.N. Wilson}
\address{Department of Physics and Astronomy, McMaster University, Hamilton,
Ontario L8S 4M1, Canada}
\author{A.M. Hallas}
\address{Department of Physics and Astronomy, McMaster University, Hamilton,
Ontario L8S 4M1, Canada}
\author{Y. Cai}
\address{Department of Physics and Astronomy, McMaster University, Hamilton,
Ontario L8S 4M1, Canada}
\author{S. Guo}
\address{Department of Physics, Zhejiang University, Hangzhou, China}
\author{Z. Gong}
\address{Department of Physics, Columbia University, New York, New York 10027,
USA}
\author{R. Sankar}
\address{Center for Condensed Matter Sciences, National Taiwan University,
Taipei 10617, Taiwan}
\address{Institute of Physics, Academia Sinica, Taipei 11529, Taiwan}
\author{F.C. Chou}
\address{Center for Condensed Matter Sciences, National Taiwan University,
Taipei 10617, Taiwan}
\author{Y.J. Uemura}
\address{Department of Physics, Columbia University, New York, New York 10027,
USA}
\author{G.M. Luke}
\address{Department of Physics and Astronomy, McMaster University, Hamilton,
Ontario L8S 4M1, Canada}
\address{Canadian Institute for Advanced Research, Toronto, Ontario M5G 1Z7,
Canada}
\begin{abstract}
We present muon spin rotation and relaxation ($\mu$SR) measurements on the
noncentrosymmetric
superconductor PbTaSe$_2$. From measurements in an applied transverse field
between $H_{c1}$ and $H_{c2}$, we extract the superfluid density as a function
of temperature in the vortex state. This data can be fit with a fully gapped
two-band model, consistent with previous evidence from ARPES, thermal
conductivity, and resistivity. Furthermore, zero field measurements show no
evidence for a time reversal symmetry breaking field greater than 0.05~G in the
superconducting state. This makes exotic fully gapped spin-triplet states
unlikely, and hence we contend that PbTaSe$_2$ is characterized by conventional
BCS s-wave superconductivity in multiple bands.
\end{abstract}
\maketitle
\section{Introduction}

Noncentrosymmetric superconductors are materials where the lack of inversion
symmetry gives rise to asymmetric spin-orbit coupling which splits otherwise
degenerate electronic bands \cite{Sigrist}. The broken symmetry removes means
that the notion of parity can no longer be used to discuss the symmetry of a
superconducting state that might emerge. This effectively
allows mixed singlet and triplet states in these materials. Mixed-parity 
superconductivity is theoretically expected to generally give rise to nodes or
partial line nodes in the gap structure 
\cite{Hayashi2006, Sigrist2007, Takimoto2009}. 

While some
noncentrosymmetric superconductors have shown evidence for line nodes,
such as CePt$_3$Si \cite{Bauer2005}, CeIrSi$_3$ \cite{Mukada2008},
Mg$_{10}$Ir$_{19}$B$_{16}$ \cite{Bonalde2009}, Mo$_3$Al$_2$C \cite{Bauer2010}
and Li$_2$Pt$_3$B \cite{Nishiyama2007}, many others display fully gapped states
\cite{Yaun2006, Isobe2016, Anand2011, Barker2015, Hillier2009, Iwamoto1998,
Yuan2006}. Multi-gap behavior has also been observed in materials like
La$_2$C$_3$ \cite{Sugawara2007}. Detailed analysis of possible microscopic
pairing mechanisms has found that either isotropic or nodal gaps can arise
depending on the anisotropy of the pairing interaction \cite{Samokhin2008};
this has also been suggested to depend on the spin-orbit coupling strength of 
the material \cite{Yaun2006}. Furthermore, even when the superconducting states
appear fully gapped, $\mu$SR measurements have found time-reversal symmetry
breaking fields in materials such as La$_7$Ir$_3$ \cite{Barker2015}
and LaNiC$_2$ \cite{Hillier2009} These varied properties make it valuable to
study additional non-centrosymmetric systems in an effort to gain a deeper
understanding of their physics.

PbTaSe$_2$ is a non-centrosymmetric material in the $P\overline{6}m2$ space
group consisting of TaSe$_2$ layers well separated by Pb interlayers, and was
recently found to be superconducting with a $T_C$ of 3.7~K \cite{Ali2014}. This
structure is similar to that of transition metal dichalcogenide (TMD) superconductors
such as TaS$_2$, IrTe$_2$ and TiSe$_2$. In these materials, the parent compound
typically hosts a charge density wave (CDW) and superconductivity emerges after
the CDW is suppressed by applied pressure, doping, or intercalation
\cite{Morosan2006, Fang2005, Yokoya2001, Yang2012, Sipos2008, Kusmartseva2009}.
Pure TaSe$_2$ has a CDW \cite{Horiba2002} that is suppressed with Pb doping
\cite{Sharafeev2015}, and so we can view PbTaSe$_2$ as a stoichiometric version
of the doped TMDs, deep inside the superconducting phase, with the novel
feature of broken centrosymmetry. ARPES measurements have also provided evidence
that the superconductivity in PbTaSe$_2$ is associated with the presence of a
nearby CDW instability \cite{Chang2016}, which further strengthens the
comparison between this compound and the doped TMDs.

The first studies of superconductivity in PbTaSe$_2$ indicated conventional
superconductivity as the magnitude of the specific heat jump is consistent with
s-wave behavior \cite{Ali2014}. However, low temperature measurements of the
upper critical field show an unconventional upward curvature \cite{Wang2015} as
a function of temperature. As
$H_{C2}(T=0)$ is still below the Pauli limit, this has been interpreted as
evidence of multi-band superconductivity rather than exotic pairing symmetry.
Furthermore, thermal conductivity measurements are consistent with fully gapped
superconductivity as there is no linear term at low temperature, and the field
dependence suggests multi-band superconductivity \cite{Wang2016}. STM
\cite{Guan2016} and ARPES \cite{Bian2016} results support this multi-band
picture as they both show multiple relevant bands near the Fermi surface.
However, despite this broad agreement, tunnel diode oscillator measurements of
the penetration depth were found to be consistent with single-band s-wave
superconductivity \cite{Pang2016}. This apparent contradiction makes it valuable
to perform complementary measurements of the penetration depth to gain
additional insight into the superconducting state of PbTaSe$_2$.

In this paper we report muon spin rotation and relaxation
($\mu$SR) measurements in the
superconducting vortex state of PbTaSe$_2$. These measurements allow us to
extract the temperature dependence of the penetration depth at two different
magnetic fields. Zero field $\mu$SR measurements also provide a sensitive test
for possible time-reversal symmetry breaking in this material. We find weak
temperature dependence to the penetration depth at low temperature that can be
characterized by fully gapped superconductivity on two bands. Furthermore, we
find no evidence for time reversal symmetry breaking.

\section{Experimental Methods}

The crystal used in this research was prepared by chemical vapor transport at
850~$^{\circ}$C using pre-reacted PbTaSe$_2$, and PbCl$_2$ as the transporting
agent. Details of the crystal growth can be found in Ref. \cite{Sankar2017}.

Muon spin rotation and relaxation ($\mu$SR) experiments were performed at the
TRIUMF laboratory
in Vancouver, Canada. We used the Pandora dilution refrigerator spectrometer on
the M15 surface-muon beam line. This instrument gives access to temperatures
between 0.03~K and 10~K with the sample mounted on a silver cold finger,
magnetic fields up to 50000~G with a superconducting magnet, and a time resolution
of 0.4~ns. The field is applied parallel to the direction of the incoming muon
beam, and we performed measurements with the muon spin rotated perpendicular to
the field direction. These experiments were performed on a thin crystal
aligned with the c-axis parallel to the muon beam. We also performed $\mu$SR
measurements in this cryostat with zero external field using copper coil
electromagnets to compensate for ambient magnetic fields. We used the $\mu$SRfit
software package to analyze the $\mu$SR data \cite{musrfit}.

Magnetometry measurements were performed at McMaster University using a Quantum
Design XL-5 MPMS with an iHelium He$^3$ cryostat insert for measurements down to
0.5~K. Magnetization curves were measured as a function of temperature
on a 3.55~mg single crystal oriented with H $\parallel$ c-axis. Alignment of the
single crystal was verified with Laue X-Ray diffraction prior to the
magnetometry and $\mu$SR measurements.

\section{Results and Discussion}

\begin{figure}[ht]
\includegraphics[width=\columnwidth]{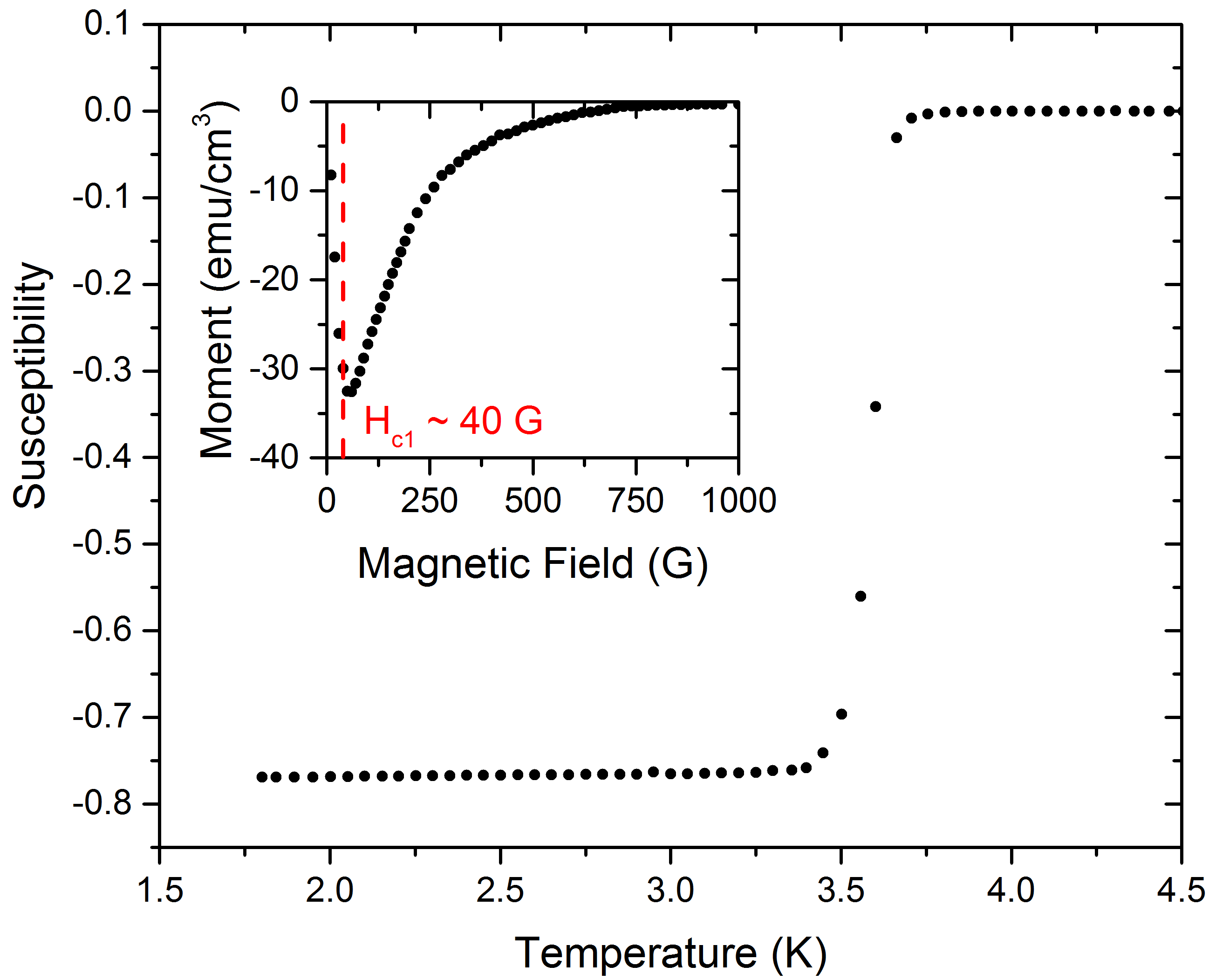}
\caption{Magnetic susceptibility measurements on a 3.55~mg single crystal of
PbTaSe$_2$ with a field of 10~G applied parallel to the c-axis after cooling in
zero field showing a superconducting transition with $T_C = 3.6$~K. (Inset)
Magnetization vs. field measurements at 0.5~K showing a lower critical field
$H_{c1} \approx 40$~G.}
\label{fig:MvT}
\end{figure}

Figure \ref{fig:MvT} shows the magnetic susceptibility versus temperature of
PbTaSe$_2$ measured in a 10~G field applied parallel to the c-axis after cooling
in zero field. The susceptibility was calculated by dividing the measured
magnetization by the applied field corrected for demagnetizing effects,
$H_{corr} = H - NM$, where $N$ is the demagnetizing factor, $H$ is the applied
field, and $M$ is the magnetization. The crystal we measured had the shape of a
thin flat plate with the field applied perpendicular to the plate.
We approximate this by
an infinitely thin flat sheet, in which case the demagnetizing factor is 1. The
susceptibility data shows diamagnetism setting in at low temperatures,
indicating a superconducting $T_C$ of 3.6 $\pm$0.1~K, in agreement with
published data \cite{Ali2014}. The sharpness of the transition, occurring over
about 0.3~K, shows that our sample is reasonably clean, as chemical or
structural disorder would broaden the transition. Furthermore, the strength of
the diamagnetic response at low temperature can be used to estimate the
superconducting volume fraction, as the susceptibility should be 1 for a pure
superconductor at low temperature. We therefore estimate the superconducting
volume fraction to be 0.77 which demonstrates that our sample is a bulk
superconductor. The small difference from one is likely from partial flux
penetration that is expected for a thin superconducting plate, or from
uncertainty in the applied field caused by flux trapping in the superconducting
magnet. The inset of Figure \ref{fig:MvT} shows the magnetic moment as a
function of applied field at 0.5~K from which we can estimate $H_{c1} \approx
40$~G and $H_{c2} \approx 1000$~G.

\begin{figure}[ht]
\includegraphics[width=\columnwidth]{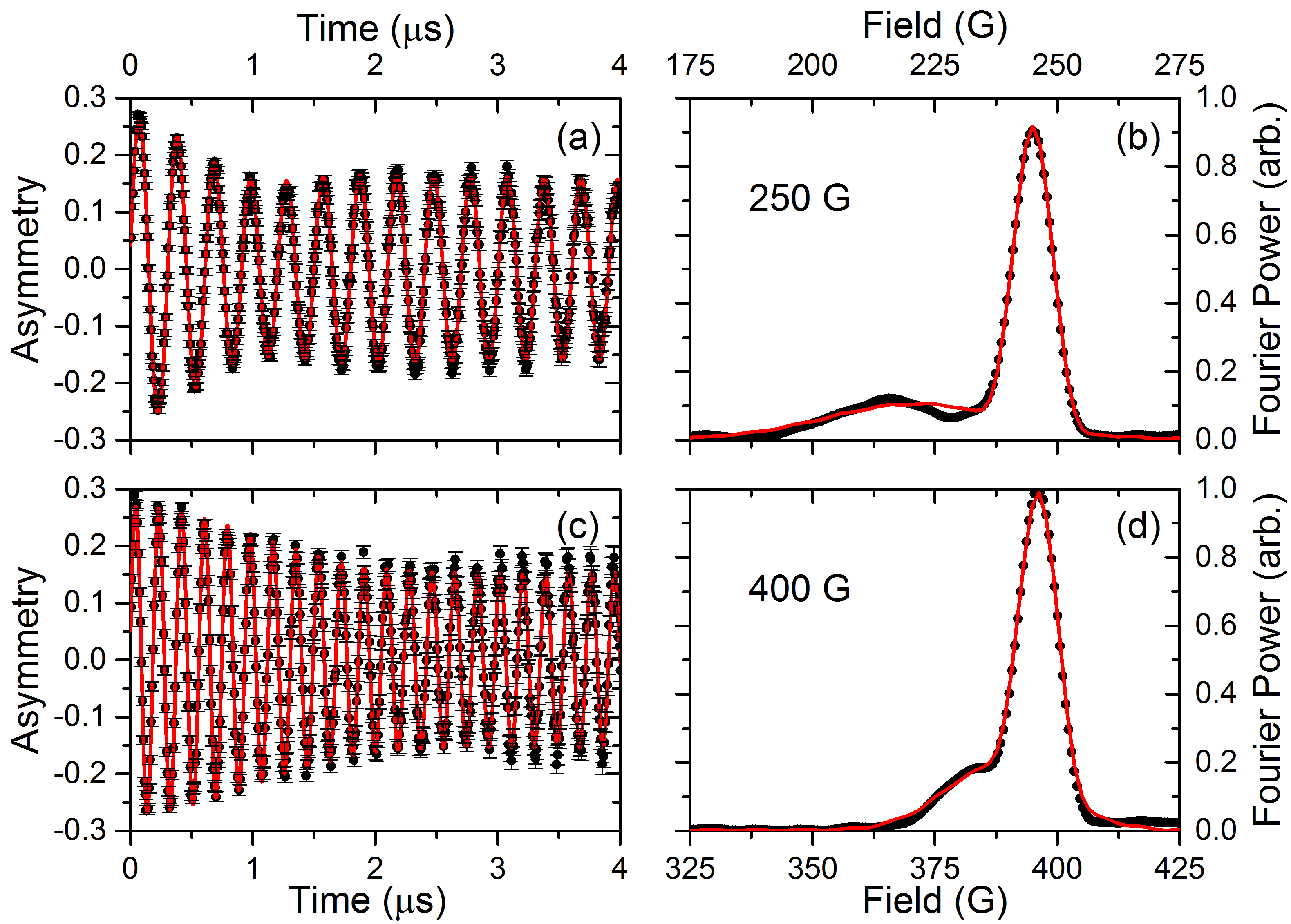}
\caption{(a) $\mu$SR asymmetry measured at 25~mK in an applied field of 250~G
$\parallel$ c-axis with the muon spins rotated $\perp$ c-axis. (b) Fourier
transform of the data shown in (a). This data shows two peaks, one coming from
muons stopping in the superconducting sample and one background peak from muons
stopping in the silver sample holder. (c) $\mu$SR asymmetry measured at 25~mK in
an applied field of 400~G $\parallel$ c-axis with the muon spins rotated $\perp$
c-axis. (d) Fourier transform of the data shown in (c). The superconducting peak
is wider and shifted closer to the background peak, indicating a larger
penetration depth and smaller diamagnetic shift at 400~G compared to 250~G.
The red lines in these figures show fits to the data following equation \ref{eq:TF}.}
\label{fig:spectra}
\end{figure}

Figure \ref{fig:spectra} (a) shows a $\mu$SR asymmetry spectrum collected at
25~mK in a 250~G $> H_{c1}$ field applied along the c-axis of our PbTaSe$_2$
sample, perpendicular to the muon spin direction. This data shows oscillations
as expected for muons precessing in an applied field, while also showing a
distinct beat in the amplitude. This demonstrates that there are two components
contributing to the asymmetry spectra, as can be seen in the Fourier transform
of the asymmetry, shown in Figure \ref{fig:spectra} (b). In this data, the large
peak just below 250~G comes from muons stopping in the silver sample holder
or non-superconducting portions of our sample,
while the peak at the lower field comes from muons stopping in the
superconducting sample. Similarly, the asymmetry spectra in Figure
\ref{fig:spectra} (c) measured at 25~mK in a field of 400~G and the
corresponding Fourier transform in Figure \ref{fig:spectra} (d) also show two
oscillating components. At this higher field the superconducting peak is
narrower and shifted closer to the silver background peak. This indicates a
larger penetration depth and a smaller diamagnetic shift at the higher field.
While the Fourier transform data is useful to make qualitative observations, it
will always contain artifacts such as peak broadening caused by the limited time
range, and hence we performed all fitting in the time domain.

Muons implanted into a type II superconductor between the lower ($H_{C1}$)
and upper ($H_{C2}$) critical fields will see the asymmetric field distribution
of the vortex state whose width is related to the London penetration depth
($\lambda$). In our sample $H_{c1} \approx 40$~G and $H_{c2} \approx 1000$~G
at low temperature, so 250~G and 400~G measurements should both be in the vortex
state. However, the small relaxation rate makes it difficult to resolve the
field distribution and we mainly see a single peak from the sample in the
Fourier transform (in addition to the background silver peak). Furthermore, the
large background peak overlaps the field region that we would expect to see the
tail of the distribution. These factors make fits to the true vortex lattice field
distribution difficult and unreliable. Instead, we fit the superconducting data
to a single Gaussian damped oscillating term, where the relaxation rate,
$\sigma_{SC}$, can be related to the penetration depth, as is commonly done for
polycrystalline samples.

\begin{figure}[ht]
\includegraphics[width=\columnwidth]{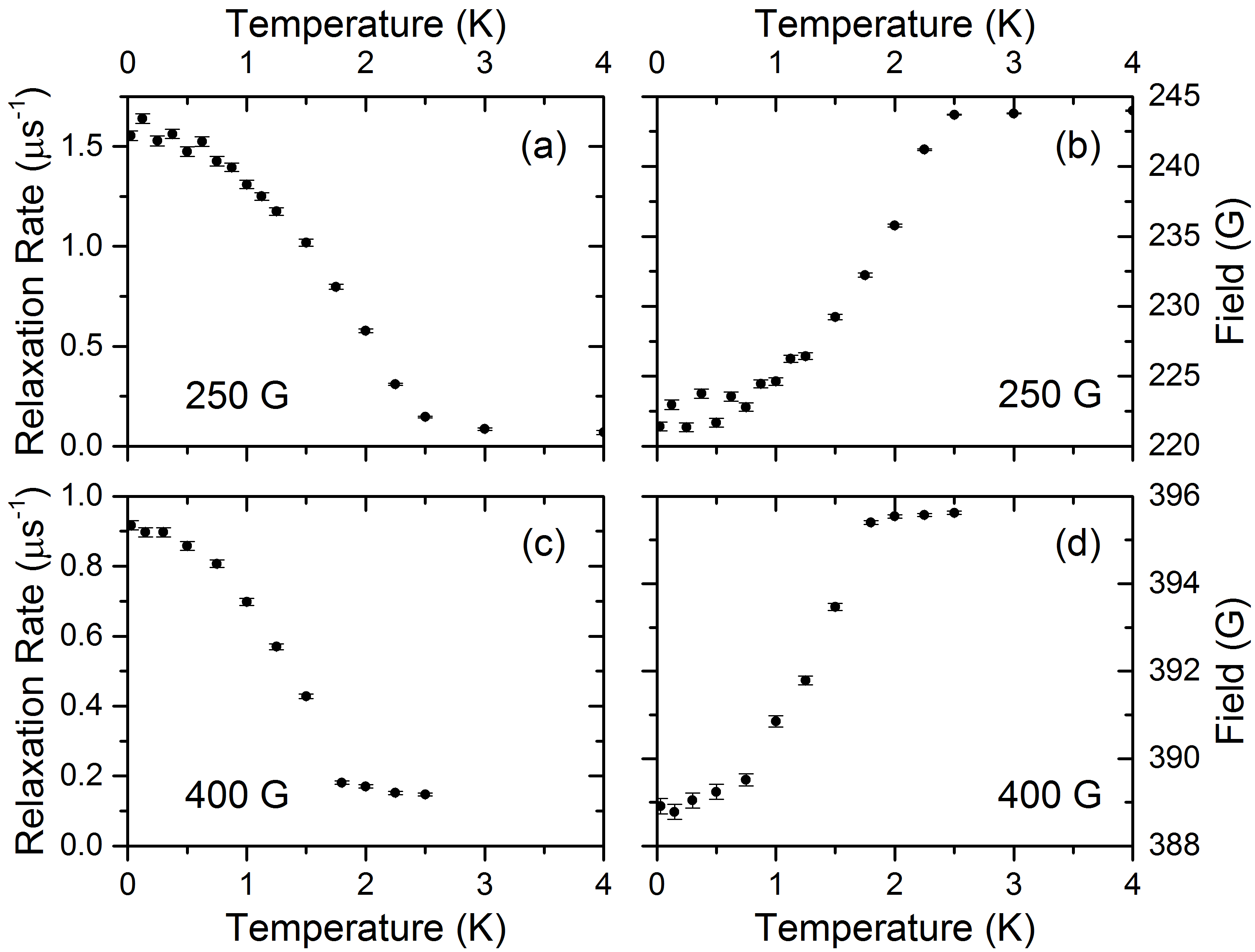}
\caption{Fit parameters extracted from fits of the $\mu$SR data measured in an
applied transverse field to equation \ref{eq:TF}. (a) $\sigma_s$ for 250~G applied
field. (b) $B_s$ for 250~G applied field. (c) $\sigma_s$ for 400~G applied
field. (d) $B_s$ for 400~G applied field.}
\label{fig:params}
\end{figure}

We fit the asymmetry data in Figures \ref{fig:spectra} (a) and (c) with equation
\ref{eq:TF} and show the fits as the red lines in Figures \ref{fig:spectra} (a)
and (c) and Fourier transformed as the red lines in Figures \ref{fig:spectra}
(b) and (d). This model has two Gaussian damped oscillating terms representing
the sample and the silver background. For the fitting, we held the total
asymmetry ($A_T$), ratio between components ($F$), silver field ($B_{Ag}$),
silver relaxation rate ($\sigma_{Ag}$), and phase ($\phi$) constant while
allowing the sample relaxation rate ($\sigma_{s}$) and field ($B_s$) to vary and
use the constant $\frac{\gamma_{\mu}}{2\pi} = 13.5538$~ kHz/G for the muon gyromagnetic
ratio. The temperature dependence of $\sigma_s$ and $B_s$ for both fields is shown in
Figure \ref{fig:params}. Figure \ref{fig:params} (a) shows an increased in 
relaxation rate setting in below 2.5~K, while Figure \ref{fig:params} (c) shows
a relaxation rate increase below 1.9~K, compared to the measured $T_C$ of 3.6~K
from Figure \ref{fig:MvT}. This is consistent with the expected suppression of
$T_C$ by an applied field for a superconductor with a relatively
low $H_{C2} \approx 1000$~G. 

From the sample relaxation rate in Figures \ref{fig:params}
(a) and (c), we determined the superconducting component of this relaxation rate
by averaging the rate above $T_C$ to determine a background rate
($\sigma_{BG}$), and then subtracting this off in quadrature from the total rate
to give $\sigma_{SC} = \sqrt{\sigma_s^2 - \sigma_{BG}^2}$.

\begin{equation}
\begin{split}
A = & A_T \left[F \cos(\gamma_{\mu}B_s t + \phi)e^{-0.5(\sigma_{s} t)^2}\right.
\\
& \left. +~ (1-F) \cos(\gamma_{\mu}B_{Ag} t + \phi)e^{-0.5(\sigma_{Ag} t)^2}
\right] \\
\end{split}
\label{eq:TF}
\end{equation}

We can relate the width of the field distribution measured by 
$\mu$SR ($\sigma_{SC} / \gamma_{\mu}$) with the penetration depth using the
relation given by equation 10 in Ref \cite{Brandt2003}. This equation gives
the variance of the magnetic field for an ideal vortex lattice, accurate for the
range of applied magnetic fields $0.25 < B / B_{c2} < 1$, which is valid for our
sample with fields down to 250~G. Expressing $\lambda$ as a function of
$\sigma_{SC}$ yields:

\begin{equation}
\lambda = \xi\sqrt{(1.94 \times
10^{-2})\frac{\phi_0}{\xi^2}(1-b)\frac{\gamma_{\mu}}{\sigma_{SC}} + 0.069}.
\label{eq:lambda}
\end{equation}
Here $\phi_0 = 2.06783 \time 10^{-15}$~Wb is the flux quantum, and $\xi$ is the
coherence length.

\begin{figure}[ht]
\includegraphics[width=\columnwidth]{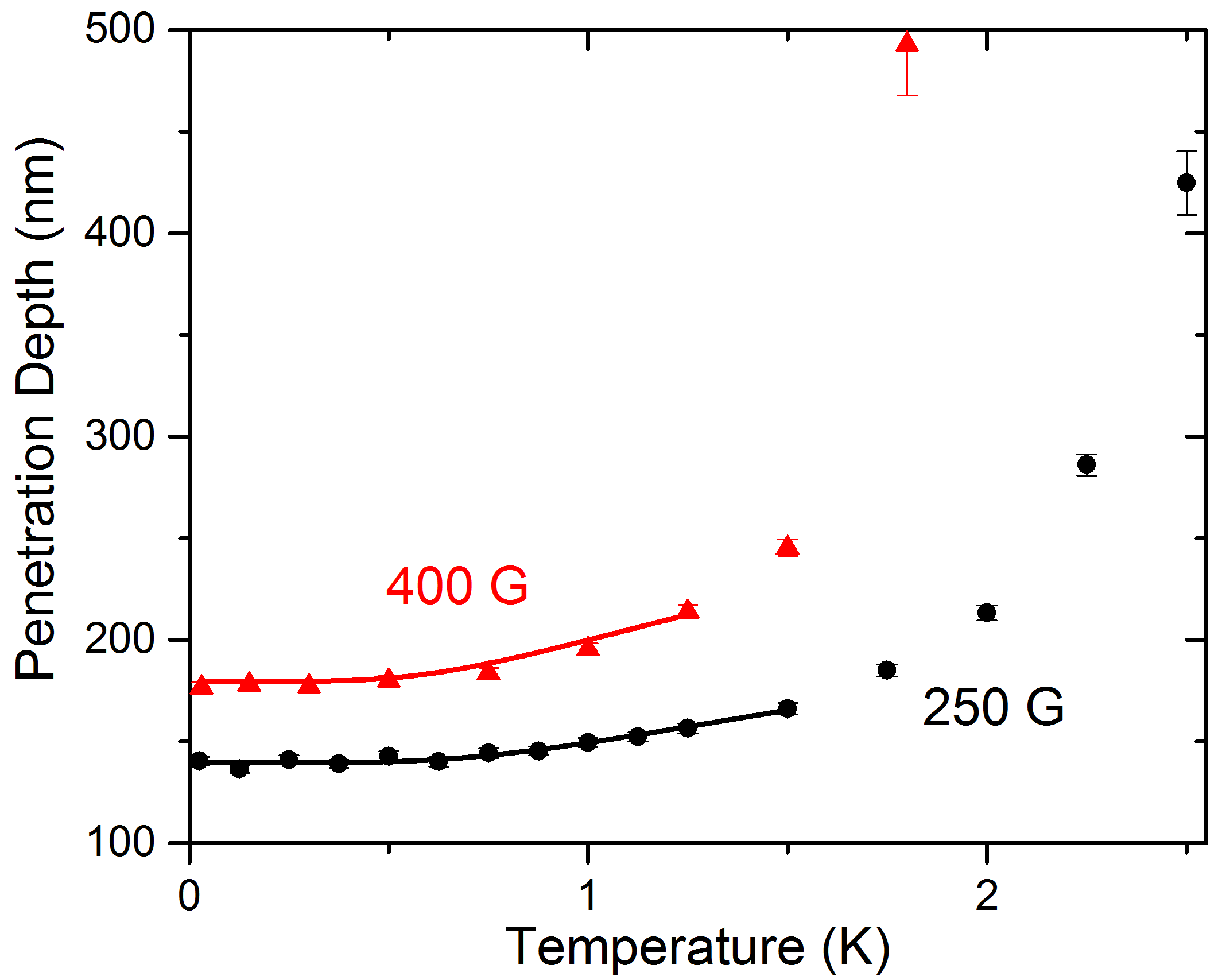}
\caption{Penetration depth calculated using equation \ref{eq:lambda} from the
$\mu$SR
data measured in 250~G (black) and 400~G (red). The solid lines show fits to the
data using equation \ref{eq:PDFIT} and show that the penetration depth scales as
expected for a fully gapped superconductor at low temperature.}
\label{fig:PD}
\end{figure}

In this equation we used $H_{c2}$ data from Ref. \cite{Wang2016} and the
relation $H_{c2} = \phi_0/(2\pi\xi^2)$ to determine $\xi$. The resultant
penetration depth is shown in Figure \ref{fig:PD}. We fit the low temperature
penetration depth, $\lambda$, with the BCS low temperature
limit\cite{Prosorov2006},

\begin{equation}
\lambda(T) = \lambda(0)\left[ 1 + \sqrt{\frac{\pi
\Delta_0}{2k_BT}}exp\left(-\frac{\Delta_0}{k_BT}\right)\right].
\label{eq:PDFIT}
\end{equation}
Here, $k_B = 8.617 \times 10^{-5} eV K^{-1}$ is Boltzmann's constant, $T$ is the
temperature, $\Delta_0$ is the zero temperature value of the gap that is
allowed to vary and $\lambda(0)$ is the zero temperature value of the
penetration depth that is also allowed to vary. These fits are shown as the
solid lines in Figure \ref{fig:PD} and show that this model fits the data well.
This suggests that the superconducting state is fully gapped without any nodes.
From these fits we extract the zero temperature penetration depth values of
$\lambda(0) = 140 \pm 1$~nm at 250~G and $\lambda(0) = 180 \pm 1$~nm at 400~G.

\begin{figure}[ht]
\includegraphics[width=\columnwidth]{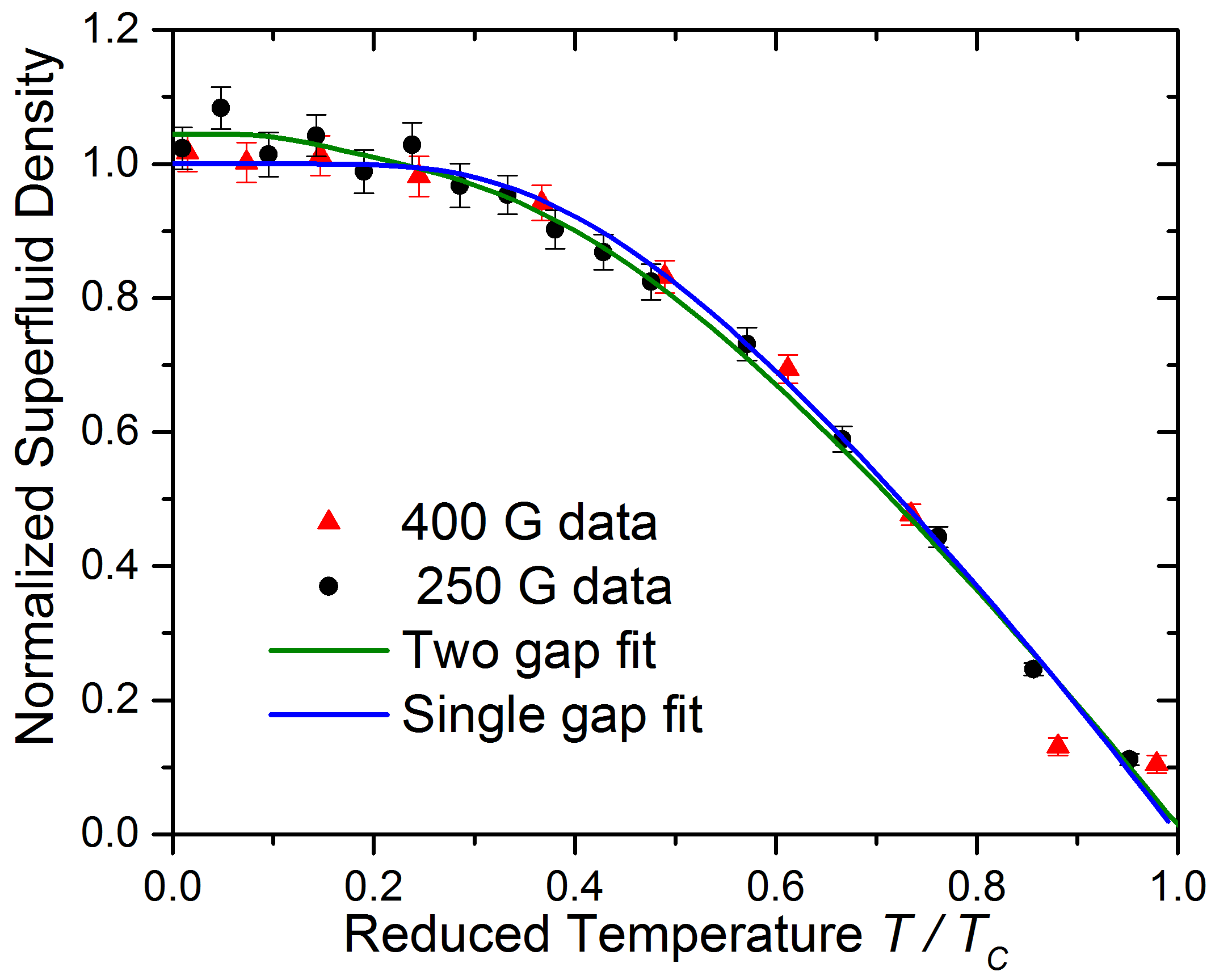}
\caption{Normalized superfluid density, $n_s(T) / n(0)$ plotted versus reduced
temperature, $T/T_C$, from the 250~G $\mu$SR data (black circles) and 400~G
$\mu$SR data (red triangles). The lines correspond to fits to a single-band
model (Equation \ref{eq:superfluid}) and a two-band model (Equation
\ref{eq:superfluid2}).}
\label{fig:SF}
\end{figure}

As a further test of the pairing symmetry of this material, we calculated the
normalized superfluid density as $n_s/n_0 = \lambda^2(0) / \lambda^2(T)$ for
both fields, and show this data in Figure \ref{fig:SF}. We then used Equation
\ref{eq:superfluid} to fit the full temperature range of this data assuming a
fully gapped superconductor.

\begin{equation}
\frac{n_s(T)}{n_0} = \left[1-2\int_{\Delta}^{\infty}dE\left(-\frac{\partial
F}{\partial E}\right)
\frac{E}{\sqrt{E^2 - \Delta^2}}\right].
\label{eq:superfluid}
\end{equation}
In this equation, $E$ is the energy difference above the Fermi energy, $F =
\frac{1}{e^{E/k_BT} + 1}$ is the Fermi function, and $\Delta$ is the gap, which
we approximate using the interpolation formula \cite{Gross1986}:

\begin{equation}
\Delta(T) = \Delta_0 \tanh\left(1.742 \sqrt{\frac{T_c}{T} - 1}\right).
\label{eq:delta}
\end{equation}

To determine the zero temperature value of the gap, $\Delta_0$, we use the BCS
weak coupling relation, $\frac{2\Delta_0}{k_BT_c} = 3.5$. This fit is shown as
the solid blue line in Figure \ref{fig:SF}, where the superfluid density at each
field, normalized to the fit zero temperature superfluid density ($n_0$), is
plotted against reduced
temperature $T/T_C$. The values $T_C(250 \mathrm{~G}) = 2.63$~K and $T_C(400
\mathrm{~G}) = 2.04$~K were determined from the fit. The fit shows reasonably
good agreement at both fields, however there is a small discrepancy at low
temperature for the 250~G data where there is an unexplained increase in the
superfluid density. A continuing increase in the superfluid density at low
temperature is commonly taken to suggest the presence of nodes in the gap
\cite{Prosorov2006}, however another possibility is multi-band superconductivity
as has notably been observed in MgB$_2$ \cite{Golubov2002}. For two uncoupled
fully gapped bands the superfluid density can be described by the sum, scaled by
some weighting factor $c$, of the contributions from the different gaps
$\Delta_1$ and $\Delta_2$ \cite{Kim2002},

\begin{equation}
\begin{split}
\frac{n_s(T)}{n_0} = &
(c)\left[1-2\int_{\Delta_1}^{\infty}dE\left(-\frac{\partial F}{\partial
E}\right)
\frac{E}{\sqrt{E^2 - \Delta_1^2}}\right] \\
+ & (1-c)\left[1-2\int_{\Delta_2}^{\infty}dE\left(-\frac{\partial F}{\partial
E}\right)
\frac{E}{\sqrt{E^2 - \Delta_2^2}}\right].
\end{split}
\label{eq:superfluid2}
\end{equation}

We used this equation to fit the 250~G data and show the fit as the solid green
line in Figure \ref{fig:SF}. This fit gives values of $c = 0.91$, $\Delta_1 =
0.399$~meV and $\Delta_2 = 0.109$~meV, suggesting a Fermi surface dominated
by a
single band, with only a small contribution coming from a second band with
smaller gap. The reduced $\chi^2$ for the two gap fit at 250~G is 1.01 compared
to 1.68 for the single band which demonstrates that it is a statistically
superior fit. However, we do not see any evidence for a continued increase in
the 400~G superfluid density with decreasing temperature as might be expected.
This could be explained by the lower gap on the
second band: with a lower gap the field required to suppress the superconducting
state is expected to be lower which would reduce the influence of multiband
behavior at 400~G compared to 250~G.

The increase at low temperature could also be explained by a 
contribution from a band with an anisotropic gap, such as a d-wave
or p-wave component. While our data alone cannot distinguish the isotropic
2-gap state from such a mixed state, comparison with data from other groups
makes the fully gapped state most likely. STM measurements show fully gapped
superconductivity \cite{Guan2016}, while thermal conductivity \cite{Wang2016}
and H$_{C2}$ \cite{Wang2015} are both consistent with multiple fully gapped 
bands. This picture also matches the theoretical expectations of Samokhin 
et al. \cite{Samokhin2008} where a noncentrosymmetric superconductor with
pairing caused by phonons should exhibit two-band nodeless superconductivity.
PbTaSe$_2$ has no surrounding magnetic phases that might promote pairing
by magnetic fluctuations, and ARPES measurements suggest 
the role of phonon stiffening in PbTaSe$_2$ compared to TaSe$_2$ to explain the
appearance of superconductivity \cite{Chang2016}. It therefore seems likely that
phonon mediated pairing is the mechanism for superconductivity in PbTaSe$_2$ 
and thus the model of Samokhin et al. with fully gapped
s-wave superconductivity would apply, consistent with our fitting.

Some pairing symmetries, notably some spin triplet p-wave states, have a
spontaneous time reversal symmetry breaking (TRSB) field in the superconducting
state. Most TRSB states are characterized by nodes in the gap which are unlikely
based on our preceding analysis. However, in high symmetry cubic
or hexagonal systems TRSB fields can appear even for fully gapped states
\cite{Agterberg1999} as has been recently reported in La$_7$Ir$_3$
\cite{Barker2015} and Re$_6$Hf \cite{Singh2014}. Such fields have in the past
been identified by $\mu SR$ measurements in zero field across the
superconducting state \cite{Luke1998, Smidman2017}. However, as the effect of
such a field is very small, significant care must be taken to minimize any stray
field at the sample position, lest a relaxation rate change caused by Meissner
expulsion of a small field be mistaken for a true time reversal symmetry
breaking field.

\begin{figure}[ht]
\includegraphics[width=\columnwidth]{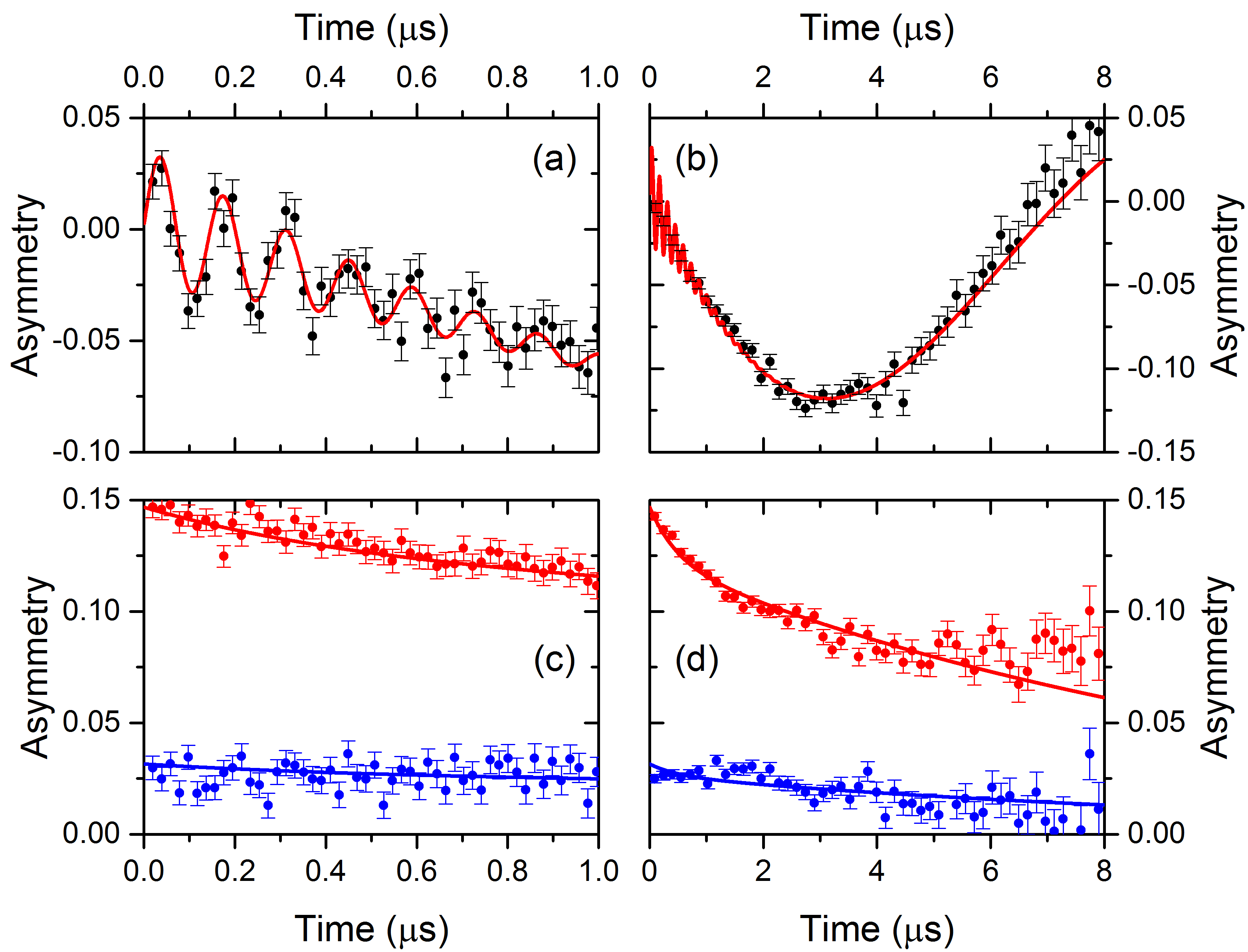}
\caption{(a) Example silicon muonium asymmetry spectra at early times in an
external field of 5~G. (b) Silicon asymmetry at long times in an external field
of 5~G. (c) Early time silicon asymmetry after zeroing the field. Red shows data
with the muon spins parallel to momentum, and blue with the muon spins rotated
perpendicular to the momentum. (d) Late time spectra corresponding to (c).}
\label{fig:zeroing}
\end{figure}

To perform these careful zero field measurements on our PbTaSe$_2$ sample
following the procedure outlined in Ref. \cite{Morris2003}, we first loaded a
piece of pure silicon in place of the sample and performed measurements at 2~K.
At this temperature, a fraction of muons landing in pure silicon
bind with electrons to form muonium, which has a gyromagnetic ratio 
$\frac{\gamma_{Mu}}{2\pi} = 1.394~MHz/G$, 103 times larger than that of a bare muon.
This gives $\mu$SR
measurements in low temperature silicon an extremely high sensitivity to small
magnetic fields. Figure \ref{fig:zeroing} (a) and (b), which show the asymmetry
spectra of silicon at 2~K in a field of 5~G, demonstrate this sensitivity.
Figure \ref{fig:zeroing} (a) shows the early time spectra with the fast
oscillations coming from muonium. Figure \ref{fig:zeroing} (b) shows the
asymmetry out to 8~ms and shows the fast oscillations coming from muonium
as well as the slower oscillations coming from bare muons. This data is fit
with two oscillating components given by equation \ref{eq:zero}, 
where the field, $B$, is the same for both components, $F_{Mu} = 0.16$
is the muonium fraction, $\lambda_{Mu}$ and $\lambda_{\mu}$ are the 
muonium and muon signal relaxation rates, and $\phi_{Mu}$ and $\phi_{\mu}$ are
phase offsets for the muonium and muon components.

\begin{equation}
\begin{split}
A = & A_T \left[F_{Mu} \cos(\gamma_{Mu}B t + \phi_{Mu})e^{-(\lambda_{Mu} t)}
\right. \\
& \left. +~ (1-F_{Mu}) \cos(\gamma_{\mu}B t + \phi_{\mu})e^{-(\lambda_{\mu} t)}
\right] \\
\end{split}
\label{eq:zero}
\end{equation}

Using this silicon sample, we zeroed the field by adjusting the currents in
copper coil electromagnets arranged in three perpendicular directions. In this
procedure, we measured spectra with the muon spin parallel and perpendicular to
the beam momentum, ensuring that the field in all directions was minimized.
Figures \ref{fig:zeroing} (c) and (d) show the silicon spectra at 2~K after
performing this zeroing procedure with the muon spins in both possible
orientations. No oscillations are observed in this data, and the field fits to a
value of $0^{+0.3}_{-0}$~G indicating that a good zero field condition was
produced.

\begin{figure}[ht]

\includegraphics[width=\columnwidth]{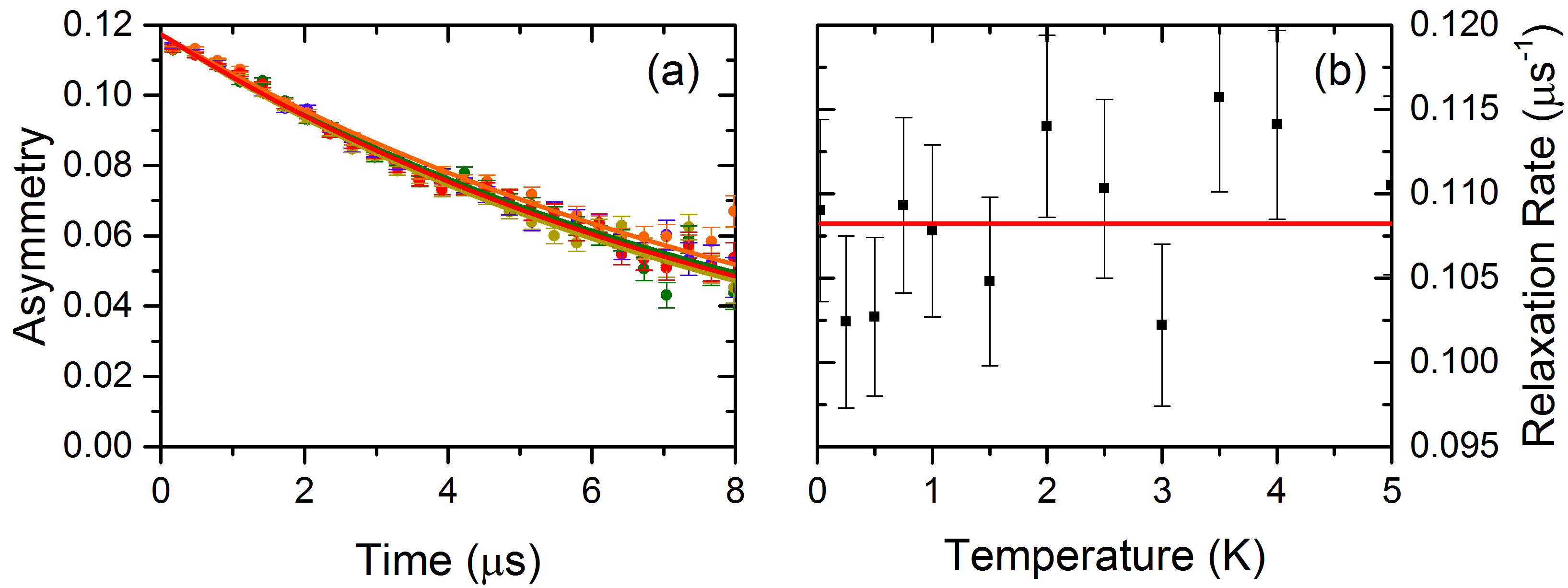}
\caption{(a) $\mu$SR asymmetry measured in zero applied field at 0.025~K (blue),
1~K (green), 2~K (yellow), 3~K (orange), and 5K (red). (b) Fit relaxation rate
for $\mu$SR data measured in zero field. The red line shows a fit to a mean
field order parameter with $T_C = 3.6$~K approximated by Equation \ref{eq:MF}.}
\label{fig:ZF}
\end{figure}

Once we arrived at this zero field condition, we re-loaded the PbTaSe$_2$ single
crystal sample in place of the silicon and performed zero field $\mu$SR
measurements with the muon spins parallel to the c-axis. Figure \ref{fig:ZF}
shows this data for representative temperatures between 0.025 and 5~K.
The asymmetry in this figure represents the signal coming only from the sample.
The background signal coming from muons stopping in the silver sample holder is
time-independent and we have allowed the baseline shift parameter ($\alpha$),
commonly used in $\mu$SR to account for varying detector efficiencies and
geometric effects, to also account for the baseline shift caused by the silver
background. This results in a total asymmetry which is comparable to the sample
component of the transverse field measurements. The relaxation of this asymmetry
in zero field will come from nuclear magnetic moments or from electronic
magnetism. As the nuclear moments will not be much affected by temperature,
and there are no known structural transitions which would modify the muon 
stopping site, any significant change in the relaxation rate with
temperature is expected to signal the onset of electronic magnetism.

Figure \ref{fig:ZF} shows no visible difference between the
asymmetry spectra at all temperatures. Fitting the data
to a single exponentially relaxing component, $A = A_T e^{-\lambda t}$, we
extract the relaxation rate ($\lambda$) and plot it in Figure \ref{fig:ZF} (b).
This figure shows no significant change in relaxation rate down to 0.025~K. To
estimate an upper limit on the internal field of the sample,
we fit Figure \ref{fig:ZF}
(b) to a mean field order parameter, approximated by Equation \ref{eq:MF}:

\begin{equation}
\lambda = c \tanh\left(1.742 \sqrt{\frac{T_c}{T} - 1}\right) + \lambda_0.
\label{eq:MF}
\end{equation}

Here $\lambda$ is the relaxation rate, $\lambda_0$ is a background relaxation
rate, $c$ is the order parameter, and $T_C$ is fixed at 3.6~K. The fit yields $c
= 0 \pm 0.004$~$\mu s^{-1}$, which gives an upper limit on the internal field of
0.05~G. Typical stray fields for time reversal symmetry breaking in triplet
superconductors are between 0.1 and 0.5~G \cite{Luke1993, Luke1998},
substantially larger than our upper limit. We therefore suggest that there is no
time reversal symmetry breaking in PbTaSe$_2$ coming from triplet pairing. In
certain non-centrosymmetric systems, smaller TRSB fields of around 0.08~G
\cite{Barker2015, Singh2014} have been reported in fully gapped superconducting
states. These fields are still larger than our fitting limit, but we cannot rule
out a slightly smaller TRSB field existing in our system.

\section{Conclusion}

Our results suggest that PbTaSe$_2$ is not purely a conventional s-wave
superconductor as the superfluid density is not flat at low temperatures. Zero
field $\mu$SR measurements find no evidence for a TRSB field, and are of
sufficient precision to rule out the field magnitudes seen in other
noncentrosymmetric and centrosymmetric superconductors. These features are
overall consistent with describing PbTaSe$_2$ as a multi-band superconductor,
with isotropic fully-gapped superconductivity existing on both bands.

\section{Acknowledgments}

We thank G.D. Morris, B.S. Hitti and D.J. Arseneau for
their assistance with the $\mu$SR measurements. Work at McMaster University was
supported by the Natural Sciences and Engineering Research Council of Canada and 
the Canadian Foundation for Innovation. M.N.W acknowledges support from the
Alexander Graham Bell Canada Graduate Scholarship program. 
The Columbia University group acknowledges support from NSF DMR-1436095 (DMREF),
NSF DMR-1610633,
OISE-0968226 (PIRE), JAEA Reimei project, and Friends of Univ. of Tokyo Inc. 
F.C.C. acknowledges support funded by the Ministry of Science and Technology
(MOST), Taiwan under Project No. 103-2119-M-002-020-MY3.

\end{document}